\documentclass[journal=jacsat,manuscript=article]{achemso}

\usepackage[version=3]{mhchem} 

\author{J\'er\^ome TRIBOLLET}
\email{tribollet@unistra.fr}
\affiliation[Strasbourg University]
{Institut de Chimie de Strasbourg, Strasbourg University, UMR 7177 (CNRS-UDS),\\
4 rue Blaise Pascal, CS 90032, F-67081 Strasbourg Cedex, 
France}

\title[An \textsf{achemso} demo]
  {Hybrid nanophotonic-nanomagnonic\\SiC-YiG quantum sensor:\\ I/ theoretical design  and properties.}

\abbreviations{EPR (Electron Paramagnetic Resonance)} 
  

\begin{document}
\begin{abstract}
Here I present the theory of a new hybrid paramagnetic-ferrimagnetic SiC-YiG quantum sensor. It is designed to allow sub-nanoscale single external spin sensitivity optically detected pulsed electron electron double resonance spectroscopy, using an X band pulsed EPR spectrometer and an optical fiber. The sensor contains one single V2 negatively charged silicon vacancy color center in 4H-SiC, whose photoluminescence is waveguided by a 4H-SiC nanophotonic structure towards an optical fiber. This V2 spin probe is created by ion implantation at a depth of few nanometers below the surface, determined by optically detected paramagnetic resonance under the strong magnetic field gradient of a YiG ferrimagnetic nanostripe located on the back-side of the nanophotonic structure. This gradient also allow the study, slice by slice at nanoscale, of the target paramagnetic sample. The fabrication process of this quantum sensor, its magnetic and optical properties, its external spins sensing properties in a structural biology context, and its integration to a standard commercially available pulsed EPR spectrometer are all presented here.
    
\end{abstract} 

\section{Introduction}
Electron paramagnetic resonance~\cite{Schweiger2001} (EPR) investigation of electron spins localized inside, at surfaces, or at interfaces of ultrathin films is highly relevant to many fields~\cite{Stesmans2017,Allouche2017,Urtizberea2018,Ternes2017,Chiesa2010}. For example, in the fields of photovoltaic~\cite{Miura2014} and photochemistry~\cite{Carbonera1998} , EPR is useful to study the spins of photo-created electron-hole pairs, their dissociation, and their eventual transport or chemical reaction occurring at some relevant interface. Also, in structural biology, it is relevant to study by EPR spin labeled proteins~\cite{Hubbel2000,Jeschke2002} introduced in model lipid bilayers membranes~\cite{Andersson2016,McCabe2013}. In the context of the development of new theranostic agents for nanomedicine, it is also relevant to study ligand-protein molecular recognition events occurring on surfaces by EPR, using for example, bifunctional spin labels~\cite{Hollas2017}. As various nanotechnologies now allow to produce biological, chemical or physical samples with nanoscale thickness, one needs to perform sensitive surface EPR. However, commercial EPR spectrometers have generally not enough sensitivity~\cite{Boero2003} 15 for studying those thin films when target spins are diluted, and, for sure, they cannot detect a single external spin. While home-made EPR experimental setups have been developed recently allowing to reach the single spin sensitivity, by optically~\cite{Degen2017,Bernardi2017,Grotz2011,Clevenson2015,Kraus2014,Widmann2015} (ODMR),  electrically~\cite{Morello2010} or mechanically~\cite{  Rugar2004} detected EPR, a powerfull upgrade strategy allowing to reach both the sub-nanoscale resolution and the single external spin sensitivity, while still using a commercially available pulsed EPR X band spectrometer, is clearly lacking and is highly relevant for most EPR users worldwide.\\\\Here I present a new methodology for state of art optically detected pulsed double electron electron spin resonance spectroscopy (OD-PELDOR), allowing SiC based sub-nanoscale single external spin sensing. This new methodology requires only a standard X band pulsed EPR spectrometer, as well as  an optical fiber and a new SiC-YiG quantum device, which can be both conveniently introduced in a standard EPR tube. The 4H SiC part of the device contains a nanophotonic structure with a single V2 negatively charged silicon vacancy color centers (spin S=3/2), introduced below the surface by low energy ion implantation and used as a quantum coherent ODMR spin probe~\cite{Kraus2014,Widmann2015,Baranov2013,Fuchs2017,Tarasenko2018} for sensing~\cite{Boero2003,Degen2017,Bernardi2017} external spins through magnetic dipolar couplings. The V2 depth below the surface can be determined by optical fiber based Optically Detected paraMagnetic Resonance (ODMR) under the strong magnetic field gradient produced by a YiG ferrimagnetic nanostripe, appropriately designed~\cite{Tribollet2014}, located nearby on the back-side of the 4H-SiC nano-photonic structure. This field gradient is also designed to allow the ODPELDOR investigation, slice by slice at sub-nanoscale, of the target paramagnetic sample located on the 4H-SiC surface. The fabrication process and the properties of this SiC-YIG quantum sensor are all presented here.\\    

\section{Principles for sub-nanoscale single external spin sensing with a V2 color center in a SiC-YiG quantum sensor.}

A single color center spin probe can sense an external spin located nearby at a nanoscale distance through its dipolar magnetic coupling with this spin (fig.1a). Since decades, the pulsed EPR experiments called DEER (Double Electron Electron Resonance) or PELDOR (Pulsed ELectron electron DOuble Resonance)~\cite{Schweiger2001,Hubbel2000,Jeschke2002,Grotz2011,Salikhov2014} allow to measure this dipolar magnetic coupling between pairs of spins A-B using pump-probe  magnetic resonance experiments with two different microwave frequencies, experiments performed over macroscopic ensemble of identical pairs of spins A-B. As this dipolar coupling scales with $\frac{1}{R^{3}_{AB}}$, $R_{AB}$ being the distance between the two spins A and B, measuring the dipolar coupling gives acces to the nanoscale distance $R_{AB}$. It was thus particularly used in structural biology for the determination of constraints on the 3D conformation of some spin labelled proteins. The situtation becomes however much more complicated for some studies like study of transmembrane proteins which are extremely difficult to obtain in large quantities and even more in 2D model lipid bio-membranes. Thus, despite the fact that 50 percent of all known small molecule drugs bind to transmembrane proteins, those proteins remains difficult to study by PELDOR spectroscopy by lack of sensitivity.\\ 

\begin{figure}[ht] 
\centering \includegraphics[width=0.6 \textwidth]{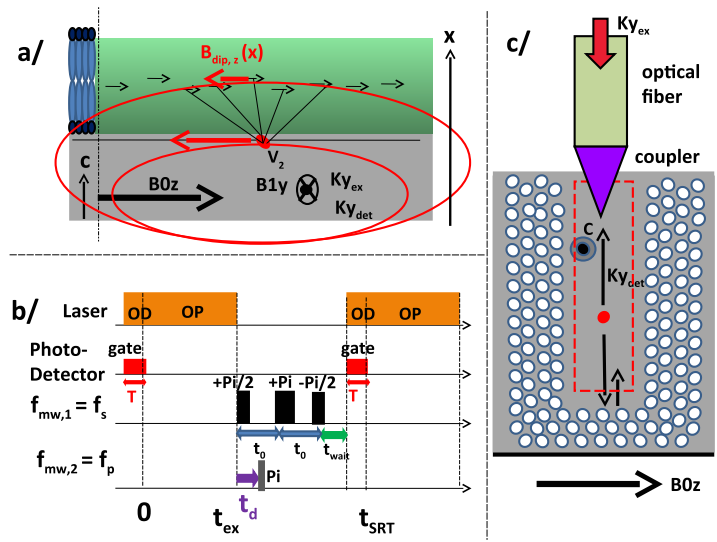}
\caption{\label{fig_01} Basic principles for external spins sensitive and nanoscale Quantum Sensing.a/ 4H-SiC-YIG sensor for ODPELDOR with a single V2 spin probe (red dot) and a plane of external target spins, like spins labels (arrows); on the left of dashed-dot line, a 2D spin labeled lipid membrane 5 nm thick, on the right, the model in green. Boz and B1y: external static and microwave magnetic fields. Bdip,z (x) is the spatially dependent dipolar magnetic field produced by the YIG ferromagnetic nanostripe located on the opposite 4H SiC surface. b/ OD-PELDOR pulse sequence for quantum sensing including synchronized optical and microwave pulses.  c/ Simplified scheme showing a top view of the SiC-YIG quantum sensor: SiC is in gray,the red dot is the V2 single color center, the dashed red rectangle is the YIG nanostripe present on the opposite surface of the SiC membrane, through which a 1D nanophotonic waveguide has been designed in a 2D photonic cristal made of air holes in SiC. The 1D optical waveguide is coupled to the optical fiber by an appropriate coupler. Ky,ex, Ky,det : excitation and PL light wavevectors.} 
\end{figure} 

 As it was previously demonstrated with a single NV color center in diamond and a home made setup~\cite{Grotz2011}, it is now possible to detect one single or several external electron spin by optically detected PELDOR spectroscopy (fig.1a). The idea is to take advantage of the spin dependent photoluminescence rate and/or of spin dependent non-radiative relaxation rates of some optically excited color centers found in wide gap semiconductors like diamond~\cite{Grotz2011}  (NV- center) or 4H-SiC~\cite{Baranov2013,Fuchs2017,Tarasenko2018} (V2 negatively charged silicon vacancy), firstly to perform optical pumping~\cite{Fischer2018} of the ground spin state of the color center, and secondly to perform an ultra-sensitive photoluminscence measurement~\cite{Kraus2014,Widmann2015} of the ground spin state population of the color center by mapping this spin state population difference onto a given photoluminescence amount. Single photon detectors in the visible or near infrared range combined with efficient optical pumping thus allow altogether to reach the single spin detection level. This is orders of magnitude better than the sensitivity of standard PELDOR spectroscopy performed using a standard pulsed EPR spectrometer operating at X band (around 10 GHz). The ODPELDOR pulse sequence (fig.1b) is thus similar to the older PELDOR pulse sequence, the main difference being the need of an initial optical pumping pulse and the need of a second optical readout pulse after the PELDOR microwave pulse sequence at two microwave frequencies, which ends by a last microwave $\frac{\pi}{2}$ pulse at probe frequency to convert the spin state coherence into a spin state population difference. The time domain ODPELDOR curve (versus td, see fig.1b for definition of td) recorded by the single V2 photoluminescence detection can be either a simple oscillation at the dipolar magnetic coupling frequency if the V2 spin is coupled to a single external spin, or a decaying curve if the single V2 spin is coupled to a spins bath (a spin ensemble), as it was previously demonstrated with a single NV- center spin probe in diamond~\cite{Grotz2011}.   
 \begin{figure*} [ht]
\centering \includegraphics[width=0.6\textwidth]{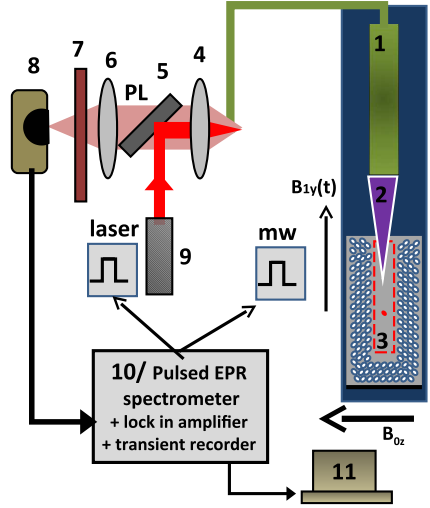} 
\caption{\label{fig_03} A laser (9) at 785 nm can conveniently excite the photoluminescence (PL) of the single V2 spin probe through a dichroic miror (5), the optical fiber (1) , the optical coupler (2) and the 1D photonic waveguide of the SiC-YIG quantum sensor (3). The PL is then collected by the same path in the reverse direction till the dichroic miror, and then it is filtered (7) before being send to the time gated sensitive photodetector (8), either a gated photomultiplier tube or a gated avalanche photodiode, having single photon counting capabilities. The photodetector signal, which is proportionnal to the photoluminescence signal, has then to be sent in a voltage form to the pulsed EPR spectrometer input channel of the transient recorder. Microwave pulses at two different frequencies are fabricated by the pulsed EPR/ELDOR spectrometer (10) and synchronized with a triger voltage pulse used to pulse the exciting laser, when necessary. For the much simpler ODMR experiments requiring a single microwave frequency, a Si photodiode can be used for PL detection and its voltage signal send to a lock in amplifier. (4) and (6) are short focal lenses or objectives. Static field B0 and microwave magnetic field B1 are perpendicular. The pulsed EPR resonator is inside a standard pulsed EPR cryostat (4K-300K). } 
\end{figure*}
As I also suggested it in a previous work on quantum computing with spins~\cite{Tribollet2014}, a strong dipolar magnetic field gradient  produced by a ferromagnetic nanostripe (see Bdip,z (x) on fig.1a) located nearby the V2 spin probe color center and its nearby external target spins, allows to encode the spatial position of each of those spins onto its paramagnetic resonance field or frequency, allowing to reach nanoscale spatial resolution with paramagnetic resonance.\\ The idea of the SiC-YIG quantum sensor presented here (fig.1c) is to provide a convenient experimental acces to this ultra-sensitive nanoscale resolution ODPELDOR spectroscopy to the many worlwide pulsed EPR users, by interfacing a standard pulsed EPR spectrometer with the SiC-YIG quantum sensor described here, using an optical fiber (fig.1c and fig.2). This 4H-SiC-YIG quantum sensor contains one single sub surface V2 negatively charged silicon vacancy color center, whose infrared photoluminescence is waveguided by a 4H-SiC nanophotonic structure (fig.1c) coupled to the optical fiber by an appropriate optical coupler~\cite{Son2018}. The SiC nanophotonic structure can be either a simple SiC nanobeam~\cite{Bracher2015,Song2018} or a 1D optical waveguide created in a 2D photonic crytsal~\cite{ Song2019,Song2011}, itself made of air holes in a thin 4H-SiC membrane~\cite{Song2019,Song2011,Cardenas2015}. A YIG nanostripe~\cite{Zhu2017} on the backside of the photonic waveguide provide a strong permanent magnetic field gradient~\cite{Tribollet2014}. Outside the pulsed EPR spectrometer, a simple photoluminescence setup using a dichroic miror and few other optics (fig.2) is used to excite and collect the photoluminescence from the V2 spin probe (at around $\lambda\:=\:915\:n\!m$ at low temperature), the light being waveguided by the optical fiber in both directions. A commercially available pulsed EPR spectrometer operating at X band is by this way upgraded into a state of art ODPELDOR spectrometer allowing single external spin sensing with nanoscale resolution. \\

\section{Fabrication process of the SiC-YiG quantum sensor}

The SiC-YIG quantum sensor proposed here is an hybrid nanophotonic-nanomagnonics device containing a 4H-SiC 1D nanophotonic waveguide and a ferrimagnetic nanostripe of YIG. Here I present one possible fabrication method of the quantum sensor, but others are possible with the advanced SiC technology.\\

\begin{figure}[ht] 
\centering \includegraphics[width=0.8 \textwidth]{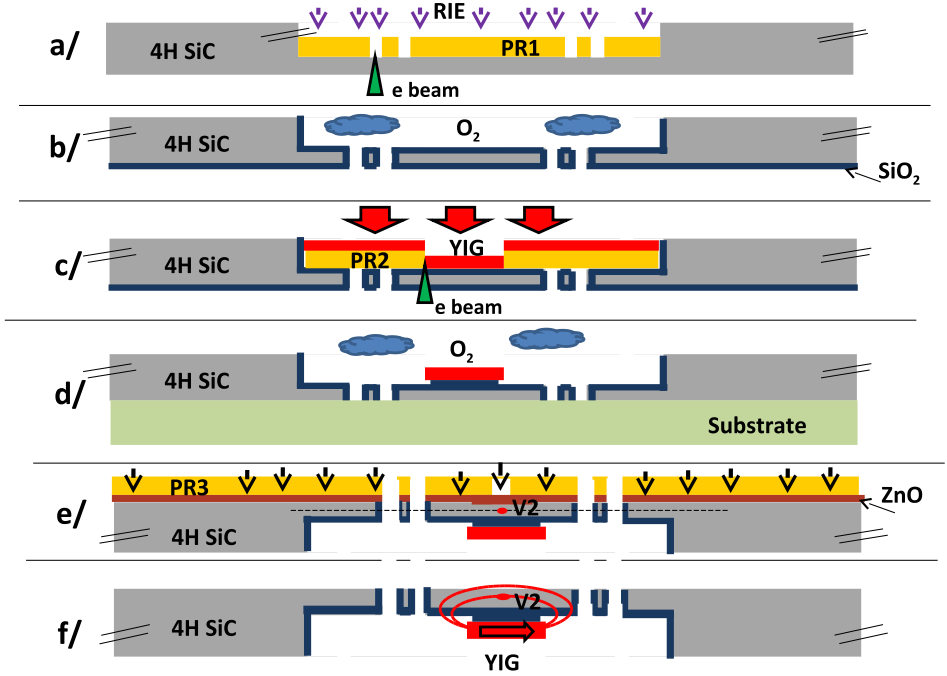}
\caption{\label{fig_03} Six main steps in the fabrication process of the 4H-SiC-YIG quantum sensor proposed. a/ SiC membrane and then 2D photonic cristal fabrication; b/ oxydation of SiC; c/ YIG nanostripe fabrication on the backside of the SiC membrane; d/ annealing under oxygen of YIG nanostripe; e/ Fabrication of a single V2 spin probe below SiC surface by ion implantation through a mask; f/ Removal of mask and final quantum device structure. See text for details.} 
\end{figure} 

It requires firstly the fabrication of a 2D photonic cristal in a thin 4H-SiC membrane, as shown on fig.3a. For this one can start with an appropriate 4H-SiC HPSI substrate with very low n type residual doping and very few residual V2 silicon vacancy and also very few other kinds of paramagnetic defects, eventually also containing a top epilayer of isotopically purified 4H-SiC to eliminate all nuclear spins from the SiC membrane to fabricate. If necessary a high temperature pre-processing annealing can be used to remove all residual V2 silicon vacancies in the starting 4H-SiC material. Then, using for example a first RIE etching step or a SICOI approach, a thin 4H-SiC membrane can be produced (around 100 nm thickness), following previous demonstrated methods~\cite{Song2019,Song2011,Cardenas2015}. If one wish to have the full SIC-YIG quantum sensor, then the SiC membrane need to be thin enough, typically between 100 nm and 300nm in order that the subsurface V2 silicon vacancy feels the optimal dipolar magnetic field gradient (see next sections). But if one wishes to produced a much simpler SiC sensor, without the YIG nanostripe, then a thicker membrane, conveniently produced by RIE can be fabricated. Then, electron beam lithography is used to fabricate an appropriate mask for the subsequent second RIE etching step allowing to produce the air holes for the photonic nanostructure. This mask has thus to reproduce the 2D photonic crystal structure wished, including the 1D photonic waveguide. This mask for RIE can be created either on the top side of SiC membrane if the membrane is thick, or on the back side of the SiC membrane if this one is very thin, using through membrane ebeam lithography~\cite{Neklyudova2017} of a photoresist (fig.3a). Then, after this RIE step and after removing the mask used, a thermal oxydation of the 4H SiC is performed (fig.3b), creating a SiO2 layer of around 10 nm. As shown on fig.3c, the single YIG nanostripe has then to be fabricated on the backside of the 4H-SiC 1D photonic waveguide. The width $W_{nanobeam}$ of this 4H-SiC 1D waveguide in the 2D photonic crystal (or the width of the alternative SiC nanobeam surrounded by air) has to be larger than the width $W_{YIG}$ of the YIG nanostripe. As the YIG nanostripe is on the backside of the device, one has to use through membrane ebeam lithography~\cite{Neklyudova2017} of a photoresist deposited on the backside(fig.3c) of the SiC membrane. Then, room temperature sputtering of YIG~\cite{Zhu2017} on SiO2 through this photoresist mask can be used to deposit YIG on the backside of the membrane. Removing the photoresist will let an amorphous YIG nanostripe deposited on the backside of the 4H-SiC 1D waveguide. A subsequent thermal annealing in oxygen (fig.3d) has then to be performed to obtain a poly-cristalline YIG nanostripe with improved spin wave resonance properties. Alternatively, direct Pulsed Laser Deposition (PLD) of YIG at high substrate temperature, followed by an appropriate backside etching of the YIG film through a mask  can be used to produce such a YIG nanostripe, but in that case one should create the YIG nanostripe on the backside of the membrane before creating the air holes of the photonic nanostructure by RIE. The mask can again in this case be defined on the backside by through membrane ebeam lithography~\cite{Neklyudova2017}.\\Then, a single V2 color center has to be introduced just few nanometers below the 4H SiC surface on the front side of the SiC membrane, at a position located vertically above the center of the YIG nanostripe itself located on the backside of the SiC membrane (fig.3e). The detailed ion implantation methods proposed to create such a single V2 spin probe in the quantum sensor are described in the appendix 1 and are inspired from a recent work reporting the fabrication of a silicon vacancies array in 4H-SiC~\cite{Wang2017}. A common key point to the various proposed statistical fabrication processes by ions implantation through a single aperture mask is the need to evaluate each quantum sensor after fabrication, because over around 100 devices fabricated in parallel, around just 5 contains an appropriate single V2 spin probe located few nm below the SiC surface, at a given depth with a precision of around +/- 1 nm. Firtsly, the second order photoluminescence auto-correlation function has to be measured to test whether a given sensor has a single V2 spin probe or not~\cite{Wang2017}. Secondly, thanks to the strong dipolar magnetic field gradient produced by the YIG nanostripe, one can determine by ODMR spectroscopy under such strong gradient, the depth of this single V2 spin probe in all sensors having a single V2. By time resolved ODMR, with an optically detected hahn echo decay measurement for example, one can also determine the T2 spin coherence time of this single V2 spin probe. Thus, the various post fabrication tests combined with an appropriate YIG nanostripe design allow an efficient selection of the quantum sensors fabricated having the desired properties (depth below the surface and T2 spin coherence time).\\ As a last remark about the whole fabrication process, one should note that in 4H-SiC, just like for diamond, surface defects exist and can induce some magnetic noise on the V2 sub-surface single spin probe. To limit this possible spin decoherence effect, a low temperature passivation treatment can be applied to the top 4H-SiC surface, like for example, a H+N plasma treatment~\cite{Liu2014} at 400 degree C, which is known to reduce its surface density of state to $6\:10^{10}\:cm^{-2}$. Thermal cooling of the quantum sensor is an alternative method to reduce or suppress the thermal fluctuations of those surface defects and thus to increase in this case the T2 of sub surface V2 spin probes.\\

\section{Design of the YIG ferrimagnetic nanostripe and magnetic properties of the SiC-YiG quantum sensor}

The quantum sensor proposed here contains a YIG nanostripe, as already said above, whose design should allow to perform ODMR test of the sub surface V2 spin probe under a strong dipolar magnetic field gradient, as well as ODPELDOR under this strong dipolar magnetic field gradient to improve the spatial resolution of ODPELDOR down to nanoscale (see fig.1 and fig.2).\\ 

\begin{figure*}[htb]
	\centering
		\includegraphics[width=0.8 \textwidth]{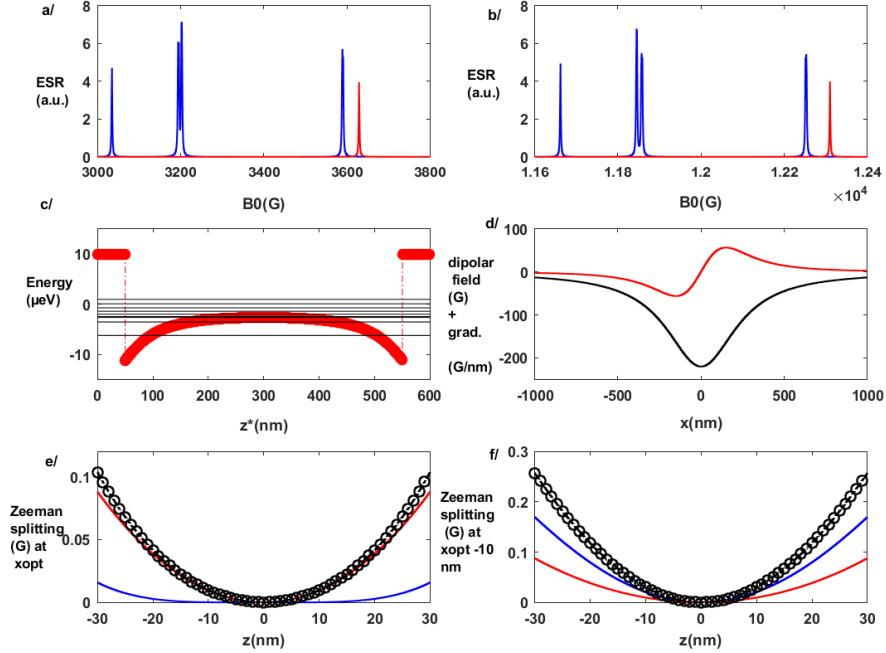}
	\caption{YIG nanostripes: width W=500 nm , thickness T=100 nm, length L=100 $\mu\:m$, Bsat=1700 G; x,y,z as in fig.1. a/ and b/ EPR spectrum at X (9.7 Hz) and Q (34 GHz) band resp., showing, in blue, YIG SWR and, in red, the shifted EPR of g=2.00 electron spins placed at xopt=150 nm above the YIG nanostripe center (x=0) (1 G linewidth for all here; SWR oscillator strength not calculated). c/ 1D eigenenergies of the spin waves along z axis; here z*=300+z; d/ z component of the dipolar magnetic field of the YIG nanostripe as a function of x (black), as well as its gradient along x (red) multiplied here by 100 for clarity. e/ and f/ Total effective Zeeman splitting at X band (dot line), expressed in Gauss  (thus divided by ($g\:\mu_{B}$), assuming g=2.00), as well as its two contributions: Bdz (to first order in blue), and Bdx (in red to second order), at xopt(e/) and at xopt +/-10 nm (f/).}
	\label{fig:5}
\end{figure*}

The figure 4 summarizes the static and dynamic magnetic properties of the YIG nanostripe located on the backside of the 1D nanophotonic waveguide. The fig.4d shows that the maximum magnetic field gradient in the x direction, perpendicular to 4H-SiC surfaces, is of around 0.5 G/nm and is obtained at a distance xopt=150 nm from the center of a given YIG nanostripe for the dimension chosen here (see legend of fig.4). That is why the SiC membrane has to have a thickness of xopt - T/2 =100 nm here, such that the V2 spin probe on the opposite face of the 4H-SiC nanophotonic waveguide feel the maximum magnetic field gradient. Increasing the width W of the YIG nanostripe will increase xopt (xopt=230 nm for W=800 nm, corresponding to a 180 nm thick 4H-SiC membrane), but it will also reduce the maximum gradient available at xopt, thus a compromise has to be found. The magnetic field gradient produced by the YIG nanostripe considered here (W=500 nm) is not rigorously one dimensional along x. However, as I previously explained in the context of quantum computing~\cite{Tribollet2014}, locally, around xopt = 150 nm here, and laterally at z=0 +/- 30 nm along z, detailed calculations clearly show (fig. 4e) that in this portion of plane above the YIG nanostripe, the dipolar magnetic field can be considered as laterally homogeneous with a precision better than 0.1 G, thus for external spins on the sensor surface. Even in the portion of plane located at around xopt - 10 nm, where the V2 spin probe could be placed, and laterally at z=0 +/- 30 nm along z, the dipolar magnetic field can be considered as laterally homogeneous with a precision better than 0.3 G (fig. 4f). In the proposed fabrication process of the V2 color center, its position is well defined at +/- 10 nm along z, the diameter of the hole in the mask used for low energy ion implantation being 20 nm. I will also assume here that it is possible to fabricate a quantum sensor device having a single isolated sub surface V2 with a spin coherence time comprised between $10\:\mu\:s$ and $50\:\mu\:s$, in principle at room temperature, and at least at a sufficiently low temperature (see appendix 2 for more details). Those spin coherence times thus correspond to a homogeneous V2 linewidth comprised between 7 mG and 36 mG, smaller than the 50 mG variation of the dipolar magnetic field which is expected with the variation of the lateral position of the V2 spin, as shown on fig.4f. This further implies, with a 0.5 G/nm gradient along x, that the V2 coordinate along the x axis, relatively to the center of the YIG nanostripe, or equivalently its depth below the 4H-SiC surface, could be determined in principle with a precision of around 1 Angstrom.\\
 
As ODPELDOR spectroscopy also allow the indirect detection of the paramagnetic resonance of an external spin through the V2 photoluminescence signal, and as this external spin feels also the field gradient of the YIG nanostripe, its paramagnetic resonance frequency (or resonant magnetic field) necessary also encode its coordinate along the x axis. Thus ODPELDOR spectroscopy performed under the strong magnetic field gradient produced by the YIG nanostripe could allow to determine the coordinate along the x axis of the external spin, relatively to the center of the YIG nanostripe, or equivalently its vertical distance above the 4H-SiC surface. At X band (around 10 GHz), at T=100K and in water-glycerol, a trityl spin label~\cite{Fielding2005} has a long longitudinal relaxation time T1= 1 ms, and also a rather long spin coherence time T2, limited by solvant proton spin diffusion and equal to around $4\:\mu\:s$. For a single trityl spin labelled transmembrane protein that would be located just few nm above the 4H-SiC surface of the SiC YIG quantum sensor, this T2 corresponds to a homogeneous linewidth of the single trityl of around 90 mG, which is slightly larger than the 0.05 G = 50 mG variation of the dipolar magnetic field which is expected with the variation of the lateral position of the spin label located betwen xopt and xopt + 10 nm, as shown on fig.4f. Thus in this case, the spatial resolution along x would be limited by the 90 mG linewidth of the trityl spin label and thus to a precision of $\frac{0.09}{0.5}\:=\:1.8$ Angstrom. This opens exciting possibilities in the context of structural biology to study the structure of spin labelled transmembrane protein reconstituted in a model lipid bilayer or lipid nanodisk, itself anchored on the 4H SiC-YIG sensor surface. This important application of this SiC-YIG ODPELDOR quantum sensor is further discussed in one of the next sections.\\ As a last remark, one can note that the YIG nanostripe design also allows to avoid the spectral overlap between the many spin wave resonance (SWR) lines of YIG and the dipolar magnetic field shifted electron paramagnetic resonance lines of the single V2 and of the single external spin label, as shown on fig.4a (at X band) and fig.4b (at Q band). As previously explained~\cite{Tribollet2014}, this is necessary to avoid the microwave driving of some spins in the YIG ferromagnetic nanostripe because this could produce some unwanted decoherence effects. Without spectral overlap between SWR and EPR lines, the residual V2 decoherence related to YIG is due to thermal fluctuations of the spins in YiG, which are expected to be negligigle here compared to other decoherence processes for the isolated sub surface V2 spin probe, following my previous estimates made for a permalloy nanostripe with a higher saturation magnetization in the context of spin based quantum computing~\cite{Tribollet2014}. Note also that the SWR mode which has the highest resonant magnetic field corresponds to a SWR mode confined on the lateral edges of the YIG nanostripe and is thus expected to have a much smaller oscillator strength compared to the quasi uniform SWR mode (FMR mode) having the largest oscillator strength and a much smaller resonant magnetic field.

\section{Design of the nanophotonic structure and optical properties of the SiC-YiG quantum sensor}

The quantum sensor proposed here contains a 4H-SiC 1D nanophotonic waveguide, as already said above, whose design should allow the efficient optical excitation of the V2 spin probe and the efficient collection of its photoluminescence, which has to be directed towards the optical fiber integrated by a coupler~\cite{Son2018} to this nanophotonic structure (see fig.1 and fig.2). This is necessary to further integrate this sensor to a widely available pulsed EPR spectrometer. For the coupling of the fiber to the nanophotonic waveguide~\cite{Son2018}, one can use either a tapered waveguide coupler, a diffraction grating based coupler, an adiabatic coupler (or evanescent coupler), or a cylindrical GRIN microlens integrated or placed at the end of the fiber. Recently, a high optical coupling efficiency above 90 percent has been demonstrated with an optical fiber coupled to a diamond nanophotonic structure~\cite{Burek2017}. An optical coupling efficiency above 25 percent has also been recently obtained between an optical fiber and a Silicon nanophotonic structure, with another approach allowing cryogenic cooling down to very low temperatures~\cite{McKenna2019}.\\ As explained in the previous section, the optimisation of the in plane homogeneity and vertical strength of the dipolar magnetic field gradient produced by the YIG nanostripe close to the top 4H-SiC surface where the V2 spin and external spins are located provides a constraint on the thickness of the SiC membrane that has to be fabricated (SiC membrane thickness = xopt - T/2, which is 100 nm for a W=500 nm YIG nanostripe width). The YIG nanostripe should also be fabricated on the backside of the 1D SiC photonic waveguide, requiring that the width of the 1D photonic waveguide be larger than the width of the YIG nanostripe. Typically, the width of the 1D photonic waveguide has to be close to an integer number of times half the wavelength of the low temperature zero phonon line of the V2 spin probe, when the 1D waveguide is a simple SiC nanobeam surrounded by air, in order to optimize photon confinement in the waveguide. With a wavelength of around 915 nm for the ZPL of V2 in 4H-SiC, this gives the following possible values of the width of such SiC nanobeam: 457 nm, 915 nm, 1372 nm,... As we also wish to have a maximum optical electric field in interaction with the sub surface V2 electric dipole, aligned along the c axis of 4H-SiC~\cite{Widmann2015,Baranov2013,Fuchs2017}, we further chose a nano-beam width either of 457 nm or 1372 nm, both having an anti-node at half such a width. For a YIG nanostripe having a width of 500 nm, one has thus to chose a SiC nanobeam width of around 1372 nm.\\ This value can be considered as a good starting value for designing also the width of another kind of 1D photonic waveguide fabricated in a 4H-SiC membrane and made of defects in a 2D photonic cristal of SiC, itself made of an array of cylindrical air holes in the SiC membrane~\cite{Song2019,Song2011,Cardenas2015}. The main optical property of this 2D photonic crystal is its optical bandgap~\cite{Song2019,Song2011}. It should be centered at the zero phonon line (ZPL) wavelength of the V2 spin probe in order that the ensemble of defects (one or several lines of defects), that are here missing air holes in such a 2D photonic crystal, create a volume for photon confinement and provide efficient 1D waveguiding properties to the V2 spin probe photoluminescence in this photonic nanostructure. The photonic band structure for TM modes propagating in the plane of a 2D photonic crystal with a triangular array of air holes in 4H-SiC can be simulated numerically~\cite{Bostan2005}. Here I focus on TM modes, that means on electromagnetic modes having an optical electric field aligned along the c axis of 4H SiC and thus having the maximal coupling with the V2 electric dipole also aligned along the c axis.\\
The results of such a numerical simulation (see Appendix 3), show that an isotorpic optical bandgap exist for TM mode if one fabricate such a nanophotnic structure with a center to center inter-hole distance equal to $a\:=\:622\:nm$ in the triangular array of air holes in SiC, the air holes having a diameter of $2\!*\!r\:=\!360\:nm$. This is quite feasible with available electron beam lithography and SiC RIE etching methods refsPCSiC. Other alternative designs based on larger but partial optical bandgap for TM modes are also possible (see discussion in Appendix 3). 

\section{Signal to noise ratio of the photoluminescence ODPELDOR signal obtained with the SiC-YiG quantum sensor}

Using the normalized DEER signal expression~\cite{Schweiger2001,Jeschke2002,Salikhov2014}, $V_{deer}$, whose value is comprised between 0 and 1, and which is directly related to the ODPELDOR experiment shown on fig.1b, considering the effect of the last additional $-\:\frac{\pi}{2}$ microwave pulse, which converts the V2 spin quantum coherence into a V2 spin state populations difference, and assuming a photon shot noise limited signal to noise ratio~\cite{Moerner2003}, one obtains the following expression for the photoluminescence signal to noise ratio $R$, in the case where ODPELDOR is obtained by off resonant optical excitation of the V2 spin probe, at 785 nm for example:  
$R\:=\:R_{opt}\:\left(1\:-\:V_{deer}\left(t_{d},\:dx,\:C_{2D\!,\:target}\right)\right)\:X$, 
and in the case where ODPELDOR is obtained by a spin state selective resonant optical excitation of the V2 spin probe, one finds $R\:=\:R_{opt}\:\left(1\:-\:V_{deer}\left(t_{d},\:dx,\:C_{2D\!,\:target}\right)\right)$, with $R_{opt}$ given by the formula: $R_{opt}\:=\:exp\left(-\frac{2\:t_{0}}{T2}\right)\!\sqrt{p_{coll}\:p_{det}\:\frac{\sigma}{A}\:\frac{P_{0}\:T}{h\:\nu}\left\langle\Phi\right\rangle}$. I used above the following notations, $\left\langle \Phi\right\rangle\:=\:\frac{\Phi_{H}\:+\:\Phi_{L}}{2}$, and  $2\:X\:=\:\frac{\Phi_{H}\:-\:\Phi_{L}}{\left\langle \Phi\right\rangle}$, taking into account the two possible different photoluminescence quantum yield for the V2 spin probe, which depend on the spin state associated to those optical transisitons (see Appendix 4). The delay td is defined on fig.1b, $C_{2D\!,\:target}$ is the 2D concentration of the target spin plane, and $dx$ is the distance between the V2 spin probe and the target spin plane. $\sigma$ is the absorption cross section of the V2 spin probe, $A$ is the area on which the optical power $P_{0}$ is sent, $h\:\nu$ is the photon energy, and $T$ is the integration time of the photoluminescence by the photodetector over one single cycle. $X$ is given above, and has a value close to 0.02 at room temperature according to previous ODMR experiments on V2 with off resonant excitation of photoluminescence~\cite{Kraus2014,Widmann2015,Baranov2013,Fuchs2017,Tarasenko2018}. $V_{deer}$ can be numerically computed using the linear approximation and shell factorization model~\cite{Jeschke2002}. This model was previously introduced for calculating the standard DEER time domain signal in the case of a three-dimensional distributions of spins. Here, this model has been adapted to take into account the bidimensional random distribution of the target external spins in their well-defined plane, parallel to the SiC sensor surface (see also next section). Assuming realistic parameters estimated in the previous sections (see Appendix 4), one finds in the case of off resonant optical excitation of the V2 spin probe, $R\:=\:90$ for a single one shot one point ODPELDOR experiment with off resonant optical excitation and a photoluminescence integration time $T\:=\:1\:\mu\!s$.\\
 
R can be off course increased in several ways. Firstly, R becomes 50 times larger when resonant optical excitation of V2 at low temperature is used, but this is more complicated in practice (see appendix 4). Secondly, using averaging over 5000 cycles of ODPELDOR experiments, R is multiplied by 70, thus reaching R= 6300 for off resonant excitation, which takes in practice around 1 second (see appendix 4). Assuming Ncycles=5000 per point and a 100 points ODPELDOR spectrum as a function of fpump (1 point each 2 MHz, 200 MHz scanned), one could obtain such a 200 MHz ODPELDOR spectrum by off resonant optical excitation (see next experimental section for an exemple of PELDOR spectrum obtained with a standard detection of EPR) in 100 s with a large signal to noise ratio R=6300, assuming negligible hardware and software delays for changing the pumping microwave frequency (otherwise, the experimental time is determined by those delays). Thus the sensitive optical, but indirect, detection of the paramagnetic resonance spectrum of Dark external spins (not photoluminescent paramagnetic centers or molecules) is possible by ODPELDOR spectroscopy. Note also that if the quantum sensor has many identical but isolated spin probes V2, $N_{V2}\:\geq\geq\:1$, located at the same depth below the SiC sensor surface, then the signal to noise ratio is in principle also enhanced by a factor $\sqrt{N_{V2}}$, but this last option for further improvement seems difficult in practice (see appendix 4).     
  
\section{ODPELDOR quantum sensing with a SiC-YiG quantum sensor applied to structural biology}

There are several ways to perform ODPELDOR quantum sensing, depending whether the quantum sensor device presented here has a YIG nanostripe or not. Also, in order to make the discussion more explicit and as structural biology studies of transmembranes proteins is expected to be an important application of this new quantum sensor, I will present the quantum sensing properties in the context of structural biology. The external spin considered here are thus spin labels, like nitroxide radicals~\cite{Schweiger2001}, gadolinium spin labels~\cite{Shah2019} and trityl radicals~\cite{Fielding2005}, whose spin hamiltonians are well known. 

\begin{figure} [ht]
\centering \includegraphics[width=1.1\textwidth]{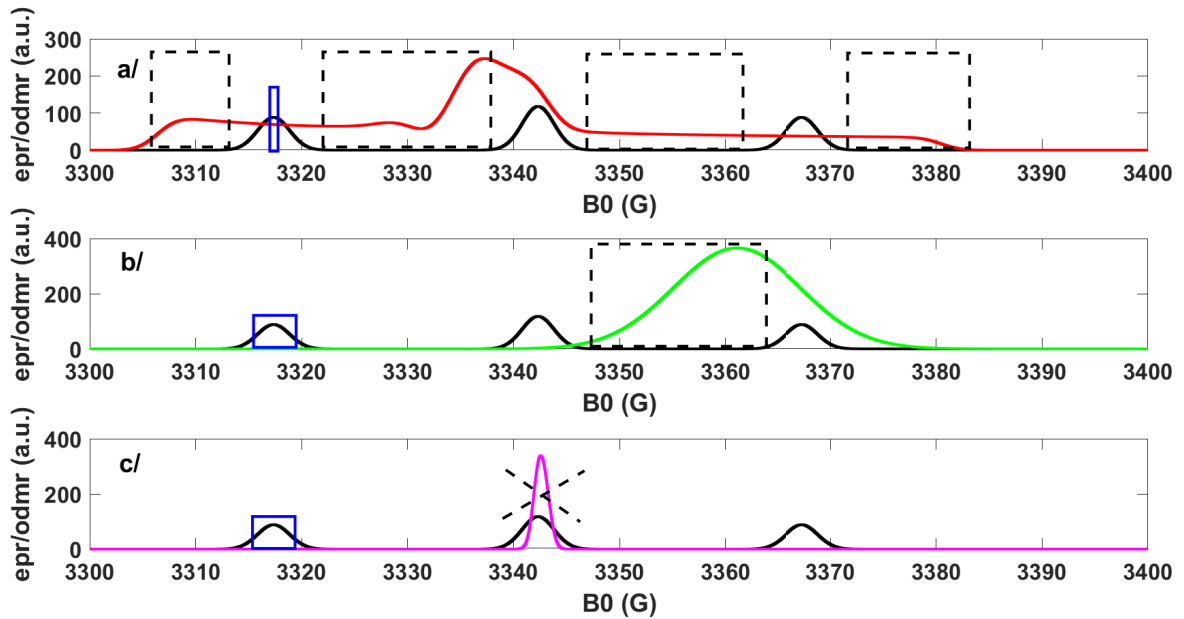} 
\caption{\label{fig_07} X band (9.369 GHz) EPR/ODMR spectrum of V2 in 4H SiC (in black) with an external magnetic field applied along the c axis of 4H-SiC, as well as a powder EPR spectrum of a/ nitroxide spin labels (red), b/ gadolinium spin labels (green), and c/ trityl spin labels (pink). Such powder spectrum is expected for spin labels in a frozen solution for example. Blue squarres for probe microwave pulse and dashed dark squarre for pump microwave pulse(s). Short pulses are spectrally broad by Fourier transform. Long pulse are spectrally selective. Multifrequency simultaneous microwave pumping is also now possible with arbitrary waveform generators (AWG).} 
\end{figure} 

So let consider firstly the quantum sensor having one single V2 spin probe and no YIG nanostripe. As already explained in the first section, the ODPELDOR pulse sequence (fig.1b) is similar to the older PELDOR pulse sequence implemented on a standard pulsed EPR spectrometer. Before performing a two microwave frequencies ODPELDOR experiment in time domain, one needs to record experimentally, or at least to numerically simulate, the EPR spectrum of both the V2 spin probe (or several ones to get all the EPR lines) and of the external spin labels. Such EPR spectrum in the first situation considered here without any YIG nanostrip is numerically simulated at X band on fig.5, using Easyspin~\cite{Stoll2006}. In this simulation I considered many V2 all having the same uniaxial magnetic anisotropy axis oriented along the C axis of 4H-SiC, also assumed here to be the direction of the externally applied magnetic field. For the spin labels, a powder spectrum was simulated with all possible orientations, providing broad spin labels EPR lines for nitroxide radicals (fig.5a) and gadolinium radicals (fig.5b), which have anisotropic g tensor and/or hyperfine coupling tensor. This is not the case for trityl spin labels, which have rather well defined magnetic parameters providing a narrow EPR line, typically of around 1 G at 70K in frozen water-glycerol~\cite{Fielding2005}. It can be seen on fig.5a and 5b, that with nitroxyde radicals and gadolinium spin labels, it is possible to excite a large part of their broad powder spectrum (dashed dark squarre), as it could be observed at low temperature, without exciting the V2 spin probe. Inversely, it is possible to excite the V2 spin probe (blue squarre) without exciting the gadolinium spin labels. In the other important case of nitroxide raidcals, which are the most common radicals used in EPR based structural biology, one can excite the V2 spin probe with a long selective resonant microwave pulse, thus minimally exciting the nitroxide spin labels. The small amount of nitroxide radicals excited by this selective excitation correspond to a very diluted external spin bath of nitroxide radicals (considering the very broad EPR spectrum associated to all the nitroxide radicals), which is thus not expected to impact significatively the spin coherence time of the V2 spin probe. One can thus perform ODPELDOR with such a SiC quantum sensor with nitroxide and gadolinium spin labels. In the case of trityl radicals however, one can see on fig.5c that the narrow EPR line of trityl raidcals will spectrally overlap one of the EPR possible transition of the V2 spin probe, the central one. Thus ODPELDOR is not possible with trityl radicals in this configuration with a 4H-SiC quantum sensor which do not have a YIG nanostripe. For the two favorable cases of nitroxide and gadolinium spin labels, the quantum sensing properties of a plane of spin labels by a single V2 spin probe are numerically simulated on fig.6 for various experimental parameters.

\begin{figure} [htb]
\centering \includegraphics[width=0.9\textwidth]{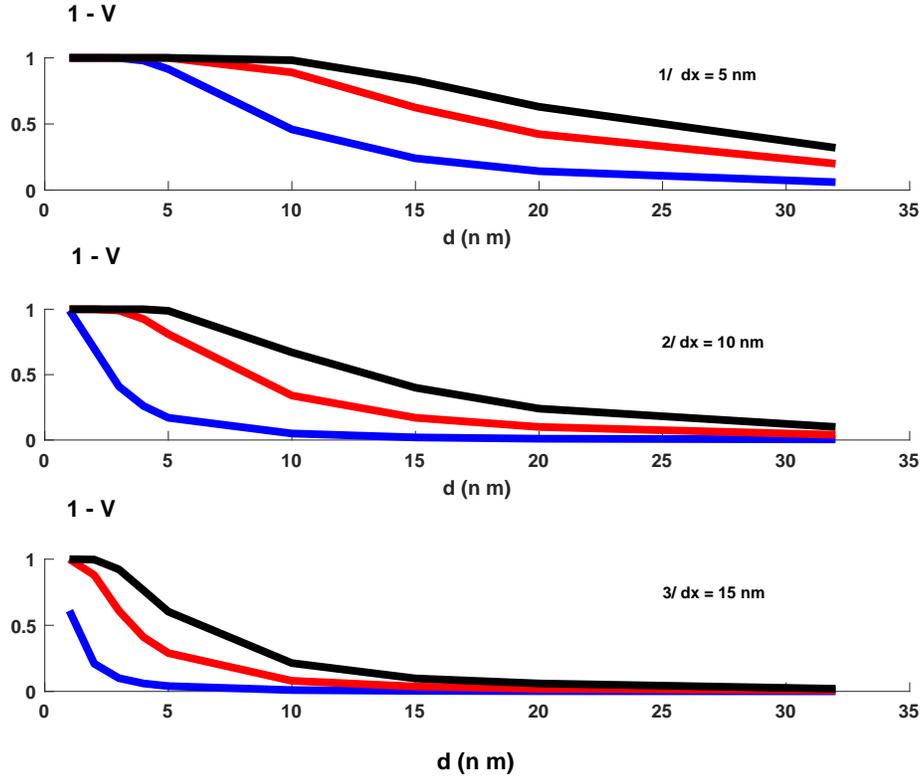} 
\caption{\label{fig_03} Dependence of the ODPELDOR normalized net signal to noise ratio (see text), 1-V, on the relative distance dx between the V2 spin probe and the plane of external spin labels (dx= 5 nm (1/) , 10 nm (2/), or 15 nm (3/), from top to bottom), and on the target spin plane concentration ($C_{2D\!,\!target}\:=\:\frac{1}{d^{2}}$, with d in nm). Dark trace is for $t_{d}\!=\!5\:\mu\!s$, red trace is for $t_{d}\!=\!3\:\mu\!s$, blue trace is for $t_{d}\!=\!1\:\mu\!s$. See fig.1 for the standard definition of td in the (OD)PELDOR sequence.} 
\end{figure} 

It is assumed on fig.6 that the spins labels are either anchored on lipids or on transmembrane proteins and located few nm above the SiC sensor surface, all in the same plane. If the ODPELDOR normalized net signal to noise ratio, 1-V, is comprised between 0.5 and 1, ODPELDOR is clearly feasible as explained in the previous section, whereas if it is between 0 and 0.5, it is much more difficult. A small inter-spin labels distance d value on fig.6 means a large 2D concentration of spin labels in the target plane to sense. One can thus see on fig.6 that ODPELDOR is much easier to perform at large 2D concentration of spin labels, at smaller dx distance between the V2 and the plane of spin labels, and it is also easier if the V2 has a long spin coherence time T2 allwoing to use a large value of the time delay td in the ODPELDOR sequence (typically $T_{2}\:\geq\:2\:t_{d}$).\\
In practice, one would perform the quantum sensing ODPELDOR experiment described here, firstly as a function of the microwave pump frequency, at fixed $t_{d}$ value, in order to detect indirectly by the V2 photoluminescence, the EPR spectrum of the external target spin labels. Then, one would perform a time domain ODPELDOR experiment, varying $t_{d}$ and using a microwave pump frequency appropriate to resonantly flip the external target spin labels. Then, the time domain decay curve obtained can be fitted with only dx and $C_{2D\:,\!target}\:=\:\frac{1}{d^{2}}$ as fitting parameters. Note that this 2D concentration is in practice often not the total 2D concentration of spin labels, but the effective concentration if one excites only one part of the target external spins spectrum, like in the case of nitroxide spin labels and gadolinium spin labels. Thus the factor associated to the proportion of the spin labels EPR spectrum effectively pumped has to be taken into account in practice. 

\begin{figure} [htb]
\centering \includegraphics[width=0.9\textwidth]{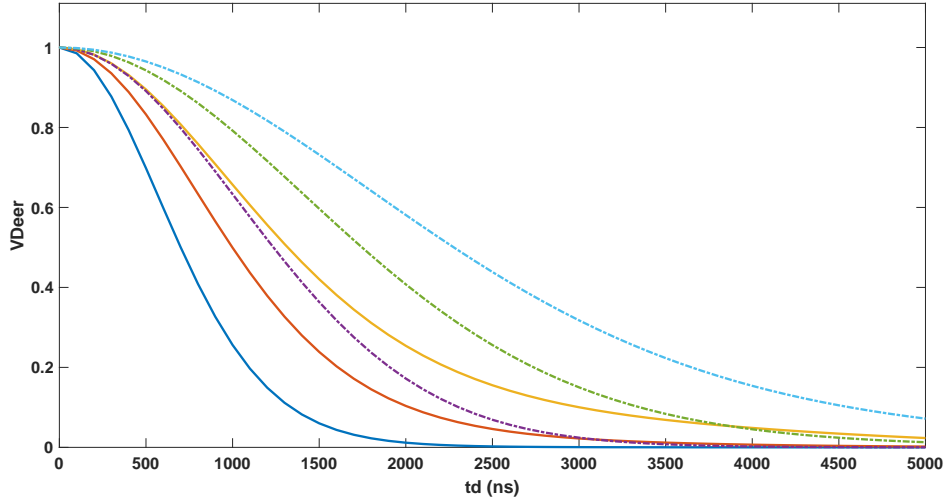} 
\caption{\label{fig_03} Time domain signal Vdeer(td), expected at dx=6nm (continuous line), and at dx=8nm (dashed dot line), each time for three different 2D external spins concentration: $C_{2D\!,\:target}\:=\:\frac{1}{\left(5\!nm\right)^{2}}$, $C_{2D\!,\:target}\:=\:\frac{1}{\left(7\!nm\right)^{2}}$, $C_{2D\!,\:target}\:=\:\frac{1}{\left(9\!nm\right)^{2}}$. The lower $C_{2D\!,\:target}$ is, the slower is the decay.} 
\end{figure} 

One can note also that state of art commercial pulsed EPR spectrometers offer the possibility to excite a given EPR spectrum over several parts by multifrequencies excitations, which is possible using an arbitrary waveform generator and which can provide a 2D concentration of spin labels effectivelly pumped close to the total 2D spin label concentration. Performing such kind of time domain ODPELDOR experiments at several spin labels concentration $C_{2D\!,\!target}\:=\:\frac{1}{d^{2}}$, it is thus possible to extract the parameter dx, directly related to the important average insertion depth of the spin labelled proteins in the model 2D biomembrane. In order to really extract this average insertion depth of the spin labelled proteins in such kind of SiC quantum sensor without a YIG nanostripe on the backside of the photonic waveguide, a first solution consist in taking advantage of the residual paramagnetic surface defects at the SiC surface as a mean to determine the exact depth of the V2 spin probe below the surface of SiC. Or, alternatively, one can fabricate on top of the SiC surface a very thin SiO2 sacrificial layer on top of which a sacrificial YIG nanostripe can be fabricated, whose magnetic field gradient can encode the depth of the V2 on its resonant frequency. Once the depth of the V2 has been determined by ODMR of the V2 spin probe under the strong magnetic field gradient of the YIG nanostripe, this YIG nanostripe and the SiO2 ultrathin layer can be removed by HF etching of the SiO2 layer. Thus simple 4H-SiC quantum sensors without a YIG nanostripe could be provided with a datasheet indicating the V2 depth below the SiC surface and its T2 spin coherence time with a free SiC surface. Note that the case of a SiC quantum sensor having an ensemble of V2 spin probe and no YIG nanostripe may have a larger signal to noise ratio but will probably have a lower spatial resolution due to the depth distribution expected by the ion implantation process (see appendix 4).\\

A second, very important kind of quantum sensor proposed here, is the SiC-YIG quantum sensor containing a single V2 spin probe and a permanent YIG nanostripe fabricated on the backside of the 1D photonic waveguide, corresponding to the case described in the fabrication process represented on fig.3. The experimental configuration here corresponds to an external magnetic field applied perpendicular to the c axis of 4H-SiC and here all spins are submitted to the strong gradient of dipolar magnetic field produced by the nearby YIG nanostripe. The simulated X band EPR/ODMR spectrum of the V2 and of the possible target spin labels, without (left) and with(right) this strong gradient is described on fig.8. This figure clearly shows that in this configuration and with the YIG nanostripe, it is now possible to perform ODPELDOR spectroscopy with a single V2 and a plane of trityl spin labels because the resonant magnetic fields of the V2 and of the trityl can be spectrally distinguished due to the different YIG induced dipolar magnetic field experienced by those spins.

\begin{figure}[ht] 
\centering \includegraphics[width=1.1 \textwidth]{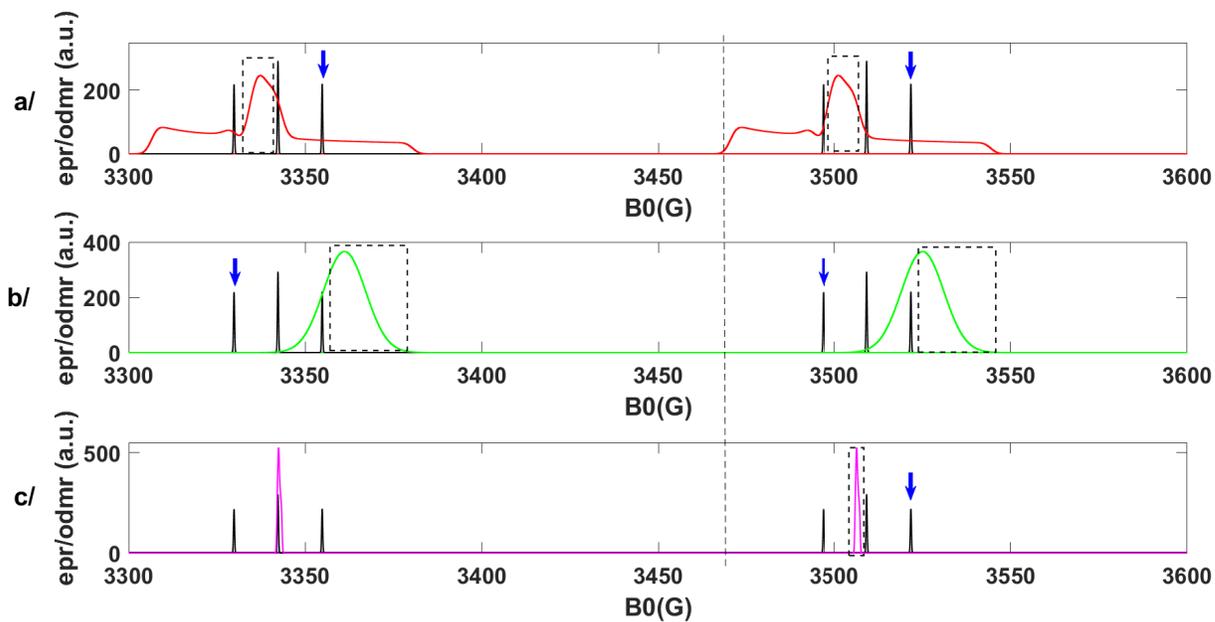}
\caption{\label{fig_10} X band (9.369 GHz) EPR/ODMR spectrum of V2 in 4H SiC (black) with an external magnetic field applied perpendicular to the c axis of 4H-SiC, as well as a powder EPR spectrum of a/ nitroxide spin labels (red), b/ gadolinium spin labels (green), and c/ trityl spin labels (pink); left side: without the YIG nanostripe gradient, right: with the YIG nanostripe gradient. Such powder spectrum is expected for spin labels in a frozen solution for example. Blue squarres for probe microwave pulse and dashed dark squarre for pump microwave pulse(s). Short pulses are spectrally broad by Fourier transform. Long pulse are spectrally selective. Multifrequency simultaneous microwave pumping is also now possible with arbitrary waveform generators (AWG).} 
\end{figure} 

The detailed magnetic properties of this SiC YIG quantum sensor have been already discussed in a previous section. Briefly, it was shown, that due to the strong dipolar magnetic field gradient produced by the YIG nanostripe, of 0.5 G per nm, and assuming $T_{2\!,\:V2}\:\geq\:10\:\mu\!s$ for a single isolated V2, and $T_{2\!,\:trityl}\:\geq\:4\:\mu\!s$ for a single isolated trityl spin label at a temperature equal or below 100 K, then one can determine the x coordinate of the V2 spin probe with a spatial resolution of around $1\:A^{\circ}$ by ODMR of the V2 under this strong gradient, and the x coordinate of the trityl with a spatial resolution of around $2\:A^{\circ}$ by ODPELDOR spectroscopy under this strong gradient, assuming a single trityl is present in the biolgical sample above the V2 spin probe. Thus ODPELDOR under the strong gradient of the YIG nanostripe of the quantum sensor allows to indirectly detect and locate with a high precision along x a single trityl, through the measurement at a given pump frequency of an oscillatory ODPELDOR time domain signal (like the one obtained in standard DEER, but here with just two spins involved).\\
Of course, it could be interesting to peform ODPELDOR spectroscopy with such an hybrid SiC-YIG quantum sensor having a doubly spin labelled single protein located just above the top SiC surface, inside a model 2D lipid bilayer or lipid nanodisc. Using the simple ODPELDOR sequence described on fig.1 but under the strong dipolar magnetic field gradient of the YIG nanostripe, then one can access, as explained just above, to the $X_{1}$ cooordinate of trityl 1 and also here to the $X_{2}$ coordinate of trityl 2, through the value of the resonant pump microwave frequencies at which one observe an oscillating ODPELDOR signal. But two other bio-structural  informations can be obtained through the fourier transformation of those two oscillating time domain ODPELDOR signal, because the oscillation period gives access to the dipolar spin-spin couplings and thus to the two relative distances between the V2 spin probe and the two trityl spin labels anchored on the target protein under study, called $R_{V\!2\:,\:Tri1}$ and $R_{V\!2\:,\:Tri2}$. Thus, ODPELDOR spectroscopy alone applied on such doubly spin labeled protein in a model lipid membrane provides already four constraints ($X_{1}$, $X_{2}$, $R_{V\!2\:,\:Tri1}$, $R_{V\!2\:,\:Tri2}$) on the 3D model of insertion or interaction of the protein with the model lipid biomembrane.

\section{Conclusion}
In this article, I have presented the theory of a new SiC-YiG hybrid quantum sensor. I have described a complete fabrication process, taking advantages of ion implantation methods and available SiC technologies, allowing to produce such a quantum sensor. The sensor has a single V2 negatively charged silicon vacancy color center introduced inside a 4H-SiC 1D nanophotonic waveguide based on defects in a 2D photonic crytsal. The waveguide is appropriately designed to waveguide efficiently the V2 spin probe photoluminescence towards an optical fiber, interfacing the pulsed EPR spectrometer with an outside standard photoluminescence setup. I showed that adding a YIG ferrimagnetic nanostripe on the backside of the SiC photonic structure is important to determine precisely the depth of the V2 spin probe below the SiC surface with angstrom resolution. I also showed that under optimal conditions, and even under off resonant optical excitation of the V2 photoluminescence, this SiC-YIG quantum sensor should allow the sub-nanoscale investigation of a single trityl spin labelled transmembrane protein by optically detected pulsed electron electron double resonance spectroscopy, with a large signal to noise ratio obtained in just one second of measurement per point, and using a simple standard X band pulsed EPR spectrometer upgraded by this quantum sensor device and a simple optical fiber. This SiC-YiG quantum sensor should thus be of great interest for all the biophysicists, chemists and physicists which are already worldwide pulsed EPR user and who wants to reach the single spin sensitivity and sub nanoscale spatial resolution offered by quantum sensing methodologies, just slightly modifying their experimental EPR setup. The next challenges are the fabrication of the proposed 4H-SiC-YIG hybrid spin-photonic-magnonic structure with state of art available SiC technology and ion implantation methods, as well as the experimental demonstration of fiber based preparatory experiments combining an external optical setup and a standard X band pulsed EPR spectrometer, both interfaced by an optical fiber.\\ 
 
\section{Acknowledgments}       

The author thanks the University of Strasbourg and the french CNRS for the reccurent research fundings.

\section{Appendices}   

Appendice 1: Ion implantation fabrication process.\\

It was recently demonstrated that low energy C+ ions ensemble implantation in 4H-SiC through an array of 65 nm diameter apertures patterned on a 300 nm thick PMMA layer using electron beam lithography allows to produce, statistically, an array of isolated single sub surface V2 silicon vacancies (see main text), having, in statistical average, a depth of 42 nm and a longitudinal straggling of about 35 nm.

\begin{figure}[ht] 
\centering \includegraphics[width=1.1 \textwidth]{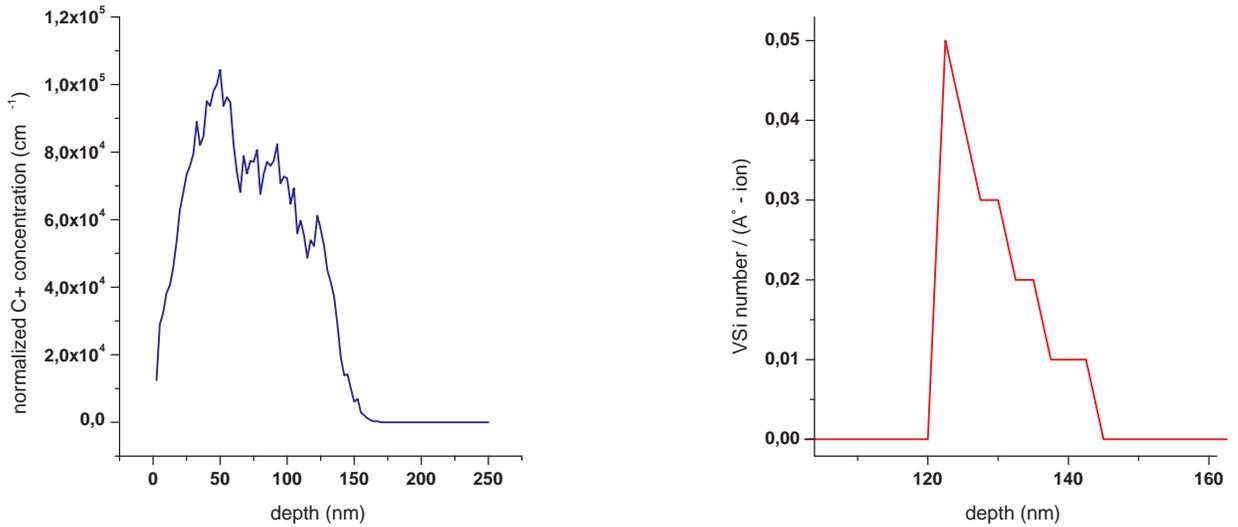}
\caption{\label{fig_10} Implanted carbon ions profile (left) as predicted by the SRIM software and assuming 30 keV C+ ions implantation (with 7 degree of tilt) in a trilayer made of ZnO (60 nm), SiO2 (60 nm) and 4H-SiC (infinity); silicon vacancy concentration profile in 4H-SiC (right).} 
\end{figure} 

 The idea here is thus to produce a mask having a single narrow aperture and thus a single access to the SiC substrate for implantation, to chose the energy of C+ ions implanted, the thickness of a sacrificial barrier layer for implantation, and also to chose the dose of C+ ions implanted, such that, in statistical average, one single V2 color center can be produced per aperture and with a control over the statistical average depth of the V2 color center. If one also targets a lateral precision better than+/- 10 nm, this corresponds to a cylindrical aperture in the mask used for ion implantation having a diameter equal to 20 nm, which is accessible by state of art electron beam lithography. The implanted carbon ions profile and the silicon vacancy concentration profile, as predicted by the SRIM software and assuming 30 keV C+ ions implantation (with 7 degree of tilt) in a trilayer made of ZnO (60 nm), SiO2 (60 nm) and SiC (infinity), are presented on fig.9. Further assuming a dose of 30 keV C+ ions implanted equal to $4.4\:10^{12}\:cm^{-2}$, as well as a decimation factor equal to 0.01 (to take into account the more realistic amount of silicon vacancies created in practice compared to the one calculted by SRIM, possibly due to interstitial-vacancy recombination processes or defects complex formation), then I find that in statistical average, one single V2 spin probe is created through an aperture of 20 nm of diameter, the other C+ ions being stopped by the 300 nm PMMA film. This also implies, in view of the Poisson Statistic expected, that if one produces in parallel 100 similar quantum sensors on the same 4H-SiC wafer, 37 quantum sensors fabricated will have exactly 1 single V2 in the aperture, 37 will have 0 V2, 18 will have 2 V2, 6 will have 3 V2,... Also, taking into account the V2 depth distribution profile of fig.9 (right), one can also calculate the probability that the V2 spin probe be contained in a thin layer of SiC of 2.5 nm thickness, as obtained by SRIM. Taking the beginning of the 4H-SiC material (absolute position 122.5 nm) as the origin of depth, I find here P(0; 2.5 nm)=0.21, P(2.5; 5 nm)=0.17,P(5; 7.5 nm)=0.14,P(7.5; 10 nm)=0.12, ... Thus one can conclude that with the fabrication process proposed here, which is statistical in essence, over 100 quantum sensor fabricated in the same way, one will obtain around $37\:*\:0.14\:=\:5$ quantum sensors having a single V2 in their aperture located above the YIG nanostripe, and with a depth comprised between 5nm and 7.5 nm. Note also that at the end of this method of implantation through a sacrificial layer, this sacrificial layer has to be removed. Here, HCl can be used to remove the ZnO sacrificial layer, while not etching the SiO2. Then HF etching can be used to remove the SiO2 sacrificial layer. Note that the SiO2 sacrificial layer below the dense ZnO layer is necessary to stop the Zn ions impacted by the C+ ions implanted, avoiding Zn contamination of the SiC top surface. As an alternative to 30 keV C+ ions implantation through a single aperture in PMMA over such trilayer (ZnO/SiO2/SiC), one could also perform a similar implantation but through a bilayer made of a thicker SiO2 sacrificial layer on SiC, targeting a maximum carbon ions and V2 concentration close to the SiO2/SiC interface and then etching the SiO2 by HF. One could also perform direct 5 keV C+ ions implantation in 4H-SiC through a single PMMA aperture of diameter 20 nm over SiC at a dose of $1\:10^{11}\:cm^{-2}$. SRIM simulation then predict in average 105 silicon vacancies per aperture, but taking into account once again the decimation factor equal to 0.01, one expect in practice in this case a single V2 spin probe per aperture. SRIm simulation (not shown here) also predict that the single silicon vacancy will be created with a maximum probability at zmax=6nm.\\

Appendice 2: homogeneous linewidth and spin coherence time for the isolated V2 in 4H-SiC.\\

An ensemble of V2 spins in 4H-SiC can already have a narrow inhomogeneous linewidth of few Gauss or even less in some previous reports. A thin epitaxially grown isotopically purified layer of SiC at the top surface of the 4H-SiC device proposed could probably further reduce this inhomogeneous linewidth. But, a single isolated V2  spin probe in 4H-SiC is expected to have an homogeneous linewidth of less than 1 G. In principle, the homogeneous linewidth should be inversely proportional to the T2 spin coherence time of the isolated V2 single spin probe.  A room temperature spin coherence time for bulk (isolated or diluted) V2 of around $50\:\mu\:s$ are often measured in bulk 4H-SiC HPSI, corresponding to a homogeneous linewidth of around 7 mG, and previous studies reported or estimated the T2 of such bulk isolated V2 to more than $300\:\mu\:s$ at room temperature. A much longer T2 is expected at low temperature at few Kelvin, because it is ultimately limited by T1 related longitudinal relaxation processes, and the T1 value of isolated V2 has been shown to exceed few minutes at few Kelvins (see main text). Indeed, using dynamical decoupling pulses sequences at low temperature, it was recently experimentally demonstrated that the effective T2 obtained using this methodology can reach $10\:m\!s$ at a temperature of few K. In practice however, for sub surface V2 created by such a fabrication process, the T2 is expected to be much smaller due to the residual surface defects density, but this remained unexplored to date and is largely dependent on the material and fabrication methodology used. A residual surface density of state of $4\:10^{10}\:cm^{-2}$, correspond to an average lattice of surface defects, separated in plane by 50 nm. In the proposed quantum sensor, the V2 spin probe could be placed around 6 nm below the 4H-SiC surface. That means that over the many SiC-YiG devices identically fabricated in parallel, there should be some of them having all the required properties, including some with a long spin coherence time due to their large distance to the nearest fluctuating surface defects. Note also that in all cases, it is in principle possible to reduce the magnetic noise due to surface defects by cooling down the sample to sufficiently low temperature, thus increasing their T1 relaxation time. \\

Appendice 3: Optical bandgap of TM modes of the 2D photonic crystal\\

Here I present a numerical simulation of the optical bandgap properties of a 2D photonic crystal, assuming the following parameters: the refractive index used are, $n_{air}\:=\:1$, $n_{SiC}\:=\:2.5$ and $r\:=\:0.29\!*\!a$, with r the radius of air holes and a the inter-center distance between neighbor air holes in the triangular lattice.
 
\begin{figure}[ht] 
\centering \includegraphics[width=0.8 \textwidth]{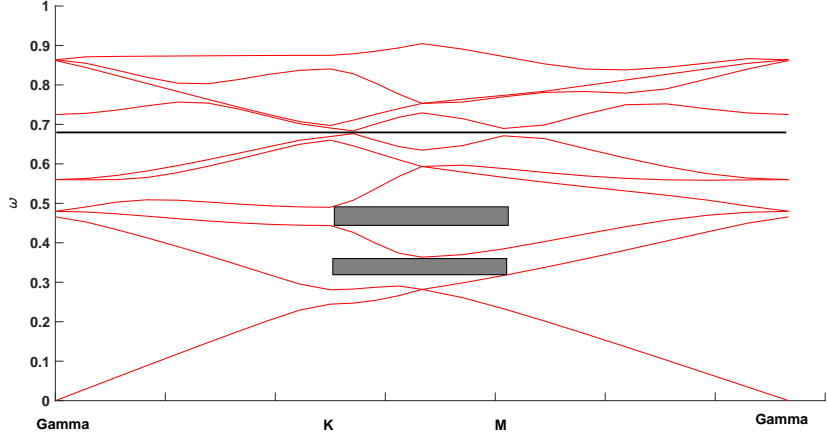}
\caption{\label{fig_6} Numerical simulation of the TM modes dispersion relation and optical bandgaps found in a triangular lattice of cylindrical air holes in 4H-SiC. The normalized frequency $\omega\:=\:\frac{a}{\lambda}$ is plotted as a function of the in plane photon wave vector considered. The horizontal axis indicates the in plane wavevector of the photon, along the direction $\Gamma\:K$, $K\:M$, and $M\:\Gamma$. Parameters: refractive index $n_{air}\:=\:1$, $n_{SiC}\:=\:2.5$ and $r\:=\:0.29\!*\!a$, with r the radius of air holes and a the inter-center distance between neighbor air holes in the triangular lattice.} 
\end{figure} 

 One can see on fig.10 that a single isotropic very narrow optical bandgap for TM modes exist for a normalized frequency $\omega\:=\:\frac{a}{\lambda}\:=\:0.680$. Thus a good design for this 2D photonic crystal consist in fabricating a triangular array of air holes in SiC, with a center to center inter-hole distance equal to $a\:=\:915\:*\:0.680\:=\:622\:nm$, the air holes having a diameter of $2\!*\!r\:=\:2\!*\!0.29\!*\!622\!=\!360\:nm$. This is quite feasible with available electron beam lithography and SiC RIE etching methods (see main text). One can also see on fig.10 that two larger but partial optical bandgap exist for TM modes propagating along the $K\:M$ direction in such a 2D photonic crystal, at normalized frequencies $\omega\:=\:\frac{a}{\lambda}\:=\:0.467$ and $\omega\:=\:\frac{a}{\lambda}\:=\:0.345$. Thus if for example, one define a line of defects in this 2D photonic crystal along the direction $\Gamma\:M$, which is perpendicular to the  $K\:M$ direction, then one expects some photon confinement in the direction perpendicular to this 1D photonic waveguide, and thus a still interesting collection efficiency for the V2 spin probe photoluminescence. More precisely, considering that a photon is emitted by the V2 with an in plane wave vector forming an angle $\theta$ with the $\Gamma\:M$ direction of this second kind of 1D photonic waveguide, then the probability of collecting this photon in the direction of the waveguide is equal to $\left(cos\left(\theta\right)\right)^{2}$. The probability that the photon is emitted in the orthogonal KM direction is $\left(sin\left(\theta\right)\right)^{2}$, and in this case it could be reabsorbed by the V2 spin probe and then reemitted in another direction more appropriate for collection by the 2D waveguide. Thus, integrating over the $\theta$ range, one finds a collection efficiency of at least 0.5 for this 1D photonic waveguide build in a 2D photonic crystal having an anisotropic optical bandgap for TM modes. This is of course less than the complete collection (collection efficiency equal to 1) which is possible with the true isotropic optical bandgap case, but it is still interesting and offer an alternative optical nanophotonic design. One possible advantage here is the larger optical partial bandgap, maybe more appropriate for room temperature operation where the V2 photoluminescence is broadened by phonon replica. In the case of a partial optical bandgap case for TM modes with $\omega\:=\:\frac{a}{\lambda}\:=\:0.467$, one thus has to fabricate a triangular array of air holes in SiC, with a center to center inter-hole distance equal to $a\:=\:915\:*\:0.467\:=\:427\:nm$, the air holes havind a diameter of $2\!*\!r\:=\:2\!*\!0.29\!*\!427\!=\!248\:nm$. This is also feasible with available electron beam lithography and SiC RIE etching methods. In the third case of a partial optical bandgap case for TM modes with $\omega\:=\:\frac{a}{\lambda}\:=\:0.345$, one thus has to fabricate a triangular array of air holes in SiC, with a center to center inter-hole distance equal to $a\:=\:915\:*\:0.345\:=\:316\:nm$, the air holes havind a diameter of $2\!*\!r\:=\:2\!*\!0.29\!*\!316\!=\!184\:nm$. This is still feasible with available electron beam lithography and SiC RIE etching methods, but maybe it is mechanically less robust.\\ 

Appendice 4: Signal to noise ratio of the ODPLEDOR experiment\\

As already said above, coupling strategies between such a 1D photonic waveguide and an optical fiber exist, with demonstrated coupling efficiencies comprised between 0.25 and 0.90. In the following, I will thus assume a V2 photoluminescence collection efficiency $p_{coll}$ comprised between 0.125=0.25*0.5 (case of partial optical bandgap) and 0.90=0.90*1 (isotropic optical bandgap), thus $p_{coll}\:=\:0.51$ in average, further assuming no loss of photons between the SiC-YIG quantum sensor and the photo-detector. The photo-detector is assumed here to have a quantum efficiency for near infrared photon detection at 915 nm of around $p_{det}\:=\:0.4$. Also, one has to note that the excitation efficiency of the photoluminescence of the V2 spin probe is proportionnal to a wavelength dependent term (related to the absorption cross section) and to the photoluminescence quantum yield, given quite generally by $\Phi\:=\:\frac{k_{rad}}{k_{rad}\:+\:k_{not-rad}}$, with $k_{rad}$ the rate of radiative recombination of the V2 spin probe from a given excited state and $k_{not-rad}$ the rate of non radiative recombination from a given excited state, which includes the inter-system crossing rate. The V2 spin probe color center has a spin $\frac{3}{2}$, and the ground and excited electronic state are both splitted by the so called zero field splitting in the ground and excited states. Thus, optical transitions noted L (like low value of ms) between GS (ms=+/-1/2) and ES (ms=+/-1/2) can be distinguished in energy from the optical transitions between GS (ms=+/-3/2) and Es (ms=+/-3/2), noted H (like high value of ms), even without application of a magnetic field. Thus, the excitation efficiency of the V2 spin probe can become spin state dependent by two effects: either by spin state selective optical excitation of the V2 spin probe at the end of the DEER sequence (after the third microwave pulse at $f_{probe}$), which is possible by energy selective resonant excitation of a given narrow optical transition by a narrow linewidth laser, or by the fact that the quantum yield is spin state dependent for the V2 spin probe, due to the spin dependent intersystem crossing rate of the V2 spin probe. I now introduce the following notations, $\left\langle \Phi\right\rangle\:=\:\frac{\Phi_{H}\:+\:\Phi_{L}}{2}$, and  $2\:X\:=\:\frac{\Phi_{H}\:-\:\Phi_{L}}{\left\langle \Phi\right\rangle}$, taking into account the two possible different photoluminescence quantum yield. Note also that it was previously shown (see main text) that the optical power necessary to obtain saturation values of optical V2 spins pumping is inversely proportional to their longitudinal spin-lattice relaxation time $T_{1}$. As $T_{1}$  increases up to several tens of second at 5K, then less than 1 mW at 780 nm spread over a 1mm*1mm square sample is sufficient at 5K for obtaining optical pumping saturation, but more at room temperature.\\With the above remarks, one can estimate the signal to noise ratio R of an ODPELDOR experiment, as described by the sequence on fig.1b, assuming $t_{d}$ is fixed and the pump frequency $f_{pump}$ vary. The photoluminescence signal $S_{pl}$ in ODPELDOR is expected to vary depending on the value of the microwave pump frequency chosen, because when the target spins are microwave manipulated on resonance, the V2 spin probe feels an accelerated spin echo decay at a given $t_{d}$ parameter value. In optimal experimental conditions, the Noise is dominated by the optical shot noise, and is given by $N_{pl}\:=\:\sqrt{S_{pl}\left(p_{B}\:=\:0\right)}$, where $p_{B}\:=\:0$ means that the microwave pump frequency is off resonant with the resonant frequency of the target external spin(s). $p_{B}\:=\:1$ means on the contrary, that the microwave pump frequency is resonant with the resonant frequency of the target external spin(s). Thus, the net signal to noise ratio R in ODPELDOR spectroscopy is given by the formula $R\:=\:\left|\frac{\left(S_{pl}\left(p_{B}\:=\:1\right)-S_{pl}\left(p_{B}\:=\:0\right)\right)}{\sqrt{S_{pl}\left(p_{B}\:=\:0\right)}}\right|$.\\
Using the normalized DEER signal expression (see main text), $V_{deer}$, whose value is comprised between 0 and 1, directly related to the ODPELDOR experiment shown on fig.1b, and considering the effect of the last additional $-\:\frac{\pi}{2}$ microwave pulse, which converts the V2 spin quantum coherence into a V2 spin state populations difference, one obtains the following expression for the signal to noise ratio $R$ in the case where ODPELDOR is obtained by off resonant optical excitation of the V2 spin probe, at 785 nm for example: 
$R\:=\:R_{opt}\:\left(1\:-\:V_{deer}\left(t_{d},\!dx,\!C_{2D\!,\!target}\right)\right)\:X$, 
and in the case where ODPELDOR is obtained by a spin state selective resonant optical excitation of the V2 spin probe (assumed on the HH optical transition), one finds $R\:=\:R_{opt}\:\left(1\:-\:V_{deer}\left(t_{d},\!dx,\!C_{2D\!,\!target}\right)\right)$, with $R_{opt}$ given by the formula: $R_{opt}\:=\:exp\left(-\frac{2\:t_{0}}{T2}\right)\!\sqrt{p_{coll}\:p_{det}\:\frac{\sigma}{A}\:\frac{P_{0}\:T}{h\:\nu}\left\langle\Phi\right\rangle}$.\\ $\sigma$ is the absorption cross section of the V2 spin probe, $A$ is the area on which the optical power $P_{0}$ is sent, $h\:\nu$ is the photon energy, and $T$ is the integration time of the photoluminescence by the photodetector over one single cycle. $X$ is given above, and has a value close to 0.02 at room temperature according to previous ODMR experiments on V2 with off resonant excitation of photoluminescence. Thus one see immediatly that R can be greatly improved by resonant spin state selective optical excitation of the photoluminescence of the V2 spin probe, on the well defined Zero phonon line (ZPL) at low temperature, but off resonant excitation is much more convenient in practice, particularly at high temperature or even at room temperature. $V_{deer}$ can be numerically computed using the linear approximation and shell factorization model (see main text). This model was previously introduced for calculating the standard DEER time domain signal in the case of a three-dimensional distributions of spins. Here, this model has been adapted to take into account the bidimensional random distribution of the target external spins in their well-defined plane, parallel to the SiC sensor surface.\\
Now, assuming a sensor operating with $t_{0}\:=\:6.25\:\mu\:s$ and  $2\:t_{0}\:=\:T_{2}\:=\:12.5\:\mu\:s$, assuming $C_{2D\!,\!target}$ is sufficiently large for a given td, such that $V_{deer}\left(t_{d},\!dx,\!C_{2D\!,\!target}\right)\:=\:0$ ie $1\:-\:V_{deer}\:=\:1$ (see main text for examples),further assuming, $\left\langle \Phi\right\rangle\:\approx\:1$, $p_{coll}\:=\:0.5$, $p_{det}\:=\:0.4$, $\frac{\sigma}{A}\:\approx\:1$, and choosing a photoluminescence integration time per ODPELDOR sequence $T\:=\:1\:\mu\!s$ for example and an optical power at 780 nm of $20\:\mu\:W$ spread over $A\:\approx\:\frac{3\:\lambda}{2}\:\frac{\lambda}{10}$, according to the previous nanophotonic structure design, one finds approximately $R_{opt}\:=\:5000$, and thus, in the case of off resonant optical excitation of the V2 spin probe, $R\:=\:5000\:\left(1\:-\:0\right)\:0.02\:\approx\:90$ for a single one shot one point ODPELDOR experiment with off resonant optical excitation.  R can be off course 50 times larger for resonant optical excitation at low temperature. The optical re-pumping time of V2 spins is short in practice with an appropriate laser, typicaly of around $150\:\mu\!s$, and much less at low temperature. The ODPELDOR microwave pulses sequence after optical initialization of V2 spins last around $20\:\mu\!s$, such that the shot repetition time of the full ODPELDOR sequence can be taken equal here to $T_{exp}\:=\:200\:\mu\!s$, meaning one can perform 5000 cycles of averaging in one second, which further improve the signal to noise ratio by a factor of $\sqrt{5000}\:\approx\:70$. As a consequence, for a SiC quantum sensor having a single V2 spin probe, one finds, assuming averaging over 5000 cycles ie over 1s of experiment in practice, a signal to noise ratio equal to $R_{1sec\:,\:off\:opt\:ex}\:=\:6300$ for off resonant excitation of the photoluminescence, while for resonant optical spin state selective excitation of photoluminescence at low temperature, one finds $R_{1sec\:,\:on\:opt\:ex}\:=\:315 000$. In both case, ODPELDOR is clearly feasible and should have a high signal to noise ratio. Even if one consider some pessimistic values like $p_{coll}\:=\:0.01$ with some photon loss betwen the quantum sensor and the photodetector, and a low efficiency detector with $p_{det}\:=\:0.04$, then one still finds for one second of averaging (5000 cycles) a signal to noise ratio of $R_{1sec\:,\:off\:opt\:ex}\:=\:286$. Assuming Nshot=5000 per point and a 100 points ODPELDOR spectrum as a function of fpump (1 point each 2 MHz, 200 MHz scanned), one could obtain such a 200 MHz ODPELDOR spectrum by off resonant optical excitation in 100 s with a signal to noise ratio R comprised between 286 and 6300, assuming negligible hardware and software delays for changing the pumping microwave frequency (otherwise, the experimental time is determined by those delays). Thus the optical, but indirect, detection of Dark external spins (not photoluminescent paramagnetic centers or molecules) is possible by ODPELDOR spectroscopy. Those estimates also assume a photon shot noise limited noise. If some photoluminescence background is present in the SiC-YIG quantum sensor device, of course the signal to noise ratio will be lower. But according to the above estimates, it seems clearly possible to tolerate some amount of photoluminescence background for performing ODPELDOR spectroscopy as long as one has sufficiently good photodetection and photon collection efficiencies, and further using averaging. Note also that the large signal to noise ratio estimated above also assume experimental parameters such that $V_{deer}\left(t_{d},\!dx,\!C_{2D\!,\!target}\right)\:=\:0$, but of course, values of $V_{deer}$ much larger, between 0.5 and 1, meaning a smaller decoherence effect of the spin bath on the V2 spin probe, could also be measured in the optimal situation of an optical shot noise limited measurement.\\                             

As a last remark, it has to be also noted that if the quantum sensor has many identical but isolated spin probes V2, $N_{V2}\:\geq\geq\:1$, located at the same depth below the SiC sensor surface, then the signal to noise ratio is in principle enhanced by a factor $\sqrt{N_{V2}}$ if it remains shot noise limited. It is however quite difficult to obtain this situation in practice by ion implantation due to the depth statistical distribution of V2 produced by this way. However, fabricating on top of the SiC surface a very thin SiO2 sacrificial layer on top of which a sacrificial YIG nanostripe can be fabricated, whose magnetic field gradient can encode the statistical distribution of the V2 depth on their resonant frequency, this statistical distribution can be determined by ODMR of the V2 ensemble under the strong magnetic field gradient, and finally the SiO2 and YIG can be removed from the top SiC surface. Then, knowing the experimentally measured statistical distribution of V2 depth, it should be possible through an appropriate fitting test function, to also extract some informations on the spatial distribution and concentration of the target external spins, with potentially a larger signal to noise ratio here but probably with a lower spatial resolution. In this kind of quantum sensor, a majority of V2 participate to the ODPELDOR experiment, but they feel different decoherence effects. The investigation of the properties of such a SiC quantum sensor having an ensemble of V2 spin probe is let for future work.\\


\section{Competing financial interests}
The author declare that he has no competing financial interests.

\end{document}